\documentclass[12pt,psfig]{article}
\usepackage{amssymb}
\usepackage{amsmath,amsthm}
\usepackage{amsfonts}
\usepackage{graphicx}
\usepackage{graphics}
\numberwithin{equation}{section}

\usepackage{hyperref}

\begin{document}
%\title{ui}\maketitle
\begin{center}\Large\textbf{Interaction 
of Dynamical Fractional Branes with Background Fields: 
Superstring Calculations}
\end{center}
\vspace{0.75cm}
\begin{center}{\large  Maryam Saidy-Sarjoubi and \large Davoud
Kamani}\end{center}
\begin{center}
\textsl{\small{Physics Department, Amirkabir University of
Technology (Tehran Polytechnic)\\
P.O.Box: 15875-4413, Tehran, Iran\\
e-mails: kamani@aut.ac.ir , mrymsaidy@aut.ac.ir\\}}
\end{center}
\vspace{0.5cm}
\begin{abstract}

We compute the boundary state corresponding to a fractional
D$p$-brane with transverse motion and internal  
background fields: Kalb-Ramond 
and a $U(1)$ gauge field. The spacetime has the 
orbifold structure 
$\mathbb{R}^{1,5} \times\mathbb{C}^{2}/\mathbb{Z}_{2}$.
The calculations are in the superstring theory. 
Using this boundary state we
shall obtain the interaction amplitude between two parallel  
moving fractional D$p$-branes. We shall extract  
behavior of the interaction amplitude for large distances of 
the branes.

\end{abstract}


\textsl{\small{Keywords}}: Boundary state; Moving fractional branes; 
Background fields; Interaction.
%$$$$$$$$$$$$$$$$$$$$$$$$$$$$$$$$$$$$$$$$$$$$$$$$$$$$$$$$$$$$
\newpage
\section{Introduction}
\hspace{0.5cm} 
The D-branes are known as objects which have critical roles in 
the string theory \cite{pol0}. They can be studied in 
various ways, for example, they may be known as 
hypersurfaces that boundaries of the string worldsheets 
can sit on them. In addition, the boundary states 
in the closed string channel  
elaborate all properties of the D-branes, such as 
their dynamics \cite{bon}, \cite{bon2}, \cite{hus}, 
\cite{bergf}, \cite{dex1}, \cite{dex2}. Precisely, 
the boundary state formalism can be used to investigate 
the interactions between the D-branes.
So far this adequate procedure for some configurations
of the D-branes, in the presence of the background fields,
for the bosonic and superstring theories has been extensively 
utilized \cite{fra}, \cite{calli}, \cite{guk}, \cite{li}, 
\cite{kit}, \cite{bach0}, \cite{dex4}, 
\cite{doc1}. Besides, the setups of the regular branes and   
systems of the fractional branes, such as BPS or 
stable non-BPS, have attracted the attention of the researchers
\cite{frac0}, \cite{dog1}, \cite{frac1}, \cite{frac2}, 
\cite{frac3}, \cite{dive}. 

Previously we have studied a system of two fractional
branes in the bosonic string theory \cite{me}.
In this paper we extend the calculations for a generalized 
setup in the superstring theory.
Thus, we consider two parallel fractional D$p$-branes 
with the Kalb-Ramond field $B_{\mu\nu}$ 
as the background field and two $U(1)$ gauge potentials 
 as the internal fields
on the worldvolumes of the branes.
The branes have velocities which are perpendicular 
to their volumes. We apply a $10$-dimensional factorizable 
spacetime with the orbifold structure, i.e.   
$\mathbb{R}^{1, 5} \times\mathbb{C}^{2}/
\mathbb{Z}_{2}$. The interaction amplitude of the branes and its 
long-range behavior will be extracted.  We shall see that our results 
considerably are novel and interesting. In fact, 
due to the fermionic degrees of 
freedom in the above-mentioned generalized 
setup, calculations are not straightforward, and drastically
we need some novel techniques. 

This paper is organized as follows. In Sec. 2, 
the boundary state of a closed superstring, 
associated with a moving fractional D$p$-brane 
with various background fields, will be constructed. 
In Sec. 3, we shall calculate the interaction
amplitude and the related long-range 
force concerning two parallel branes in the twisted NS-NS 
and R-R sectors of the superstring theory. 
Section 4 is devoted to the conclusions.
%%%%%%%%%%%%%%%%%%%%%%%%%%%%%%%%%%%%%%%%%%%%%%%%%%%%%%%%%%%%%%%%
\section{The boundary state}
\hspace{0.5cm}
A fractional brane of our system is stuck at the fixed-points 
of the noncompact orbifold
$\mathbb{C}^{2}/\mathbb{Z}_{2}$, where 
the complex coordinates of $\mathbb{C}^{2}$ are constructed from 
$\{x^a|a= 6,7,8,9\}$, which the $\mathbb{Z}_2$-group acts on them.
The fixed-points are located at the hyperplane $x^a = 0$.

In this section we extract the boundary state of a moving 
fractional D$p$-brane with background fields through 
the sigma-model action of closed string.
Let us have a glance on the  
bosonic part of the string sigma-model that we use
\begin{eqnarray}
S=&-&\frac{1}{4\pi\alpha'}\int_{\Sigma} d^2\sigma
\left(\sqrt{-h} h^{ab}G_{\mu\nu} \partial_{a}X^\mu
\partial_b X^{\nu} +\epsilon^{ab}B_{\mu\nu}\partial_a X^{\mu}
\partial_b X^{\nu}\right)
\nonumber\\
&+&\frac{1}{2\pi\alpha'}\int_{\partial\Sigma}d\sigma A_{\alpha}
\partial_{\sigma}X^{\alpha}~,
\end{eqnarray}
where the set $\{x^\alpha | \alpha = 0,1, \cdot\cdot\cdot,p\}$
shows the brane directions,
$\Sigma$ indicates the worldsheet of a closed string and
$\partial\Sigma $ is its boundary. The variables $h_{ab}$ and
$G_{\mu\nu}$ are the metrics of the string 
worldsheet and the $10$-dimensional spacetime, which will 
be considered flat metrics, e.g. $G_{\mu\nu}=\eta_{\mu\nu}=
{\rm diag}(-1,1,\cdot\cdot\cdot,1)$. 
For the $U(1)$ gauge potential we consider the gauge 
$A_{\alpha}=-\frac{1}{2}F_{\alpha \beta }X^{\beta}$, where 
the field strength $F_{\alpha \beta}$ is constant. 
In addition, for the next purposes,
we consider the constant Kalb-Ramond field $B_{\mu\nu}$.

\subsection{The boundary state equations}

\subsubsection{The bosonic part }

Variation of the action with respect to $X^\mu(\sigma,\tau)$
gives the equations of motion and the boundary state equations.
Imposing a boost on the resulted 
boundary state equations, along the 
non-orbifoldy perpendicular 
directions to the brane, we obtain the boundary state equations 
of a moving fractional brane \cite{me}, 
\begin{eqnarray}
&~&[\gamma\partial_{\tau}(X^0-v^i X^i)
+\gamma \mathcal{F}^{0}_{~~{\bar
\alpha}} \partial_\sigma
X^{{\bar \alpha}}]_{\tau=0}|B_X\rangle=0~,
\nonumber\\
&~&[\partial_{\tau}X^{\bar{\alpha}}
+\gamma^2\mathcal{F}^{\bar{\alpha}}_{~~0}
\partial_\sigma (X^0-v^i X^i)
+\mathcal{F}^{\bar{\alpha}}_{~~{\bar \beta}}
\partial_\sigma X^{\bar \beta}]_{\tau=0}|B_X\rangle=0~,
\nonumber\\
&~&[\gamma(X^i-v^i X^{0})-\gamma y^i]_{\tau=0}|B_X\rangle=0~,
\nonumber\\
&~&[X^a-y^a]_{\tau=0}|B_X\rangle=0~,\label{2.3}
\end{eqnarray}
where $\gamma = \sqrt{1-v^iv^i}$,
the set $\{x^{\bar \alpha}|{\bar \alpha}=1,\cdot\cdot\cdot,p\}$ 
shows the spatial directions
of the brane, and $\{x^i|i=p+1,\ldots,5\}$ 
indicates the directions perpendicular
to its worldvolume, except the orbifoldy directions. 
The parameters $\{y^i|i=p+1,\ldots,5\} 
\bigcup \{y^a| a=6,7,8,9\}$ represent 
the initial location of the brane.
Since the branes are stuck at the
orbifold fixed-points we have $\{y^{a}=0| a=6,7,8,9\}$.
For simplifying the equations we assumed that
the mixed 
elements $B^\alpha_{~~i}$ and $B^\alpha_{~~a}$ to be zero.
The total field strength is defined by 
$\mathcal{F}_{\alpha\beta}=F_{\alpha\beta} -B_{\alpha\beta}$. 

Note that the dimension of the orbifold crucially puts a bound on the 
dimension of the brane. In order to have a
velocity perpendicular to the brane, 
the maximum spatial dimension of the brane is 4, i.e. $p \leq 4$.

\subsubsection{The fermionic part}

To find the boundary state equations for the worldsheet fermions, 
we accurately perform the worldsheet
supersymmetry on the bosonic boundary equations (\ref{2.3}). That is,
we use following replacements
\begin{eqnarray}
\partial_{+}X^\mu(\sigma,\tau)&\rightarrow&
-i\eta\psi_{+}^\mu(\sigma,\tau)~,
\nonumber\\
\partial_{-}X^\mu(\sigma,\tau)&\rightarrow&-\psi_{-}^\mu(\sigma,\tau)~,
\label{jk}
\end{eqnarray}
where $\eta=\pm1$ is introduced for the 
GSO projection of the boundary state.
To obtain the replacement for $X^\mu$ in terms of 
the fermionic coordinates, Eq. (\ref{jk}) induces    
\begin{equation}
X^\mu(\sigma,\tau)\rightarrow\sum_{t}\frac{1}{2t}
\left(i\psi_t^\mu~e^{-2it(\tau-\sigma)}
+\eta \tilde{\psi}_t^\mu~ e^{-2it(\tau+\sigma)}\right)~.\label{ki}
\end{equation}
 
It is well known that the spectrum of the string in the presence of the 
orbifoldy directions is a linear combination of two sectors: 
the states which are invariant under the orbifold projection, 
which are called the untwisted sector, and the states of the twisted 
sector which are living at the orbifold fixed-points
\cite{frac2}. Therefore, in general, the boundary state 
of a fractional brane in an orbifold theory can be written as 
\begin{eqnarray}
|B\rangle=\mathcal{N}^{\rm U}\left(|B\rangle^{\rm U}_{\rm NS}+\epsilon_1
|B\rangle^{\rm U}_{R}\right)
+\mathcal{N}^{\rm T}\left(|B\rangle^{\rm T}_{\rm NS}
+\epsilon_2|B\rangle^{\rm T}_{\rm R}\right)~,\label{dft}
\end{eqnarray}
where `U' and `T' indicate the untwisted and twisted sectors, 
$\mathcal{N}^{\rm U}$ and $\mathcal{N}^{\rm T}$ 
are normalization constants, 
$\epsilon_1$ and $\epsilon_2$ are two R-R charges ($\epsilon=+1,-1$  
stands for brane and anti-brane, respectively). 
Eq. (\ref{dft}) has been written 
in this form to save the generality,
which admits one to consider various 
configurations of the branes (brane-antibrane) in the both 
twisted and untwisted sectors. 

We are not going to perform calculations for the complete 
boundary state (\ref{dft}). This is due to the fact that
the untwisted states $|B\rangle^{\rm U}_{\rm NS , R}$ 
represent the usual boundary states 
with the standard GSO-projection which 
have been vastly studied for 
various setups \cite{dex1}, \cite{dex2},  
\cite{fra}, \cite{calli}, \cite{guk}, \cite{li}, 
\cite{kit}, \cite{bach0}, \cite{dex4}, 
\cite{doc1}. Thus, we  shall not repeat 
the calculations for this sector, instead 
we shall just concentrate on  
the twisted states $|B\rangle^{\rm T}_{\rm NS , R}$ 
in the next sections. 

For the boundary state equations in the twisted 
part, we have to use the fermionic mode expansions of 
the twisted sector, which are given by
\begin{eqnarray}
\psi_{+}^\mu=\sum_{t}\tilde{\psi}_{t}^\mu 
e^{-2it(\tau + \sigma)}~,~~~~~~~~~~~~
\psi_{-}^\mu=\sum_{t}\psi_{t}^\mu e^{-2it(\tau-\sigma)}~,
\end{eqnarray}
where in the twisted R-R sector there are 
\begin{eqnarray}
&~& ~\psi_{t}^\lambda ~\text{and}~ \tilde{\psi}_{t}^\lambda, 
~~ t\in \mathbb{Z} ,\nonumber\\
&~&\psi_{r}^a ~\text{and} ~\tilde{\psi}_{r}^a,~~ r\in 
\mathbb{Z}+\frac{1}{2},\nonumber
\end{eqnarray}
where $\{\lambda |\lambda=0,1,2,3,4, 5\}$ stand for the  
non-orbifoldy directions and 
$\{a|a=6,7,8,9\}$ represent the 
orbifoldy directions. In the twisted NS-NS sector we have
\begin{eqnarray}
&~& ~\psi_{t}^\lambda ~\text{and}~ \tilde{\psi}_{t}^\lambda, 
~~ t\in \mathbb{Z}+\frac{1}{2},\nonumber\\
&~&\psi_{r}^a ~\text{and} ~\tilde{\psi}_{r}^a,~~ r\in \mathbb{Z}.
\nonumber
\end{eqnarray}

By replacing Eqs. (\ref{jk}) and (\ref{ki}) into Eq.  
(\ref{2.3}), and applying the above mode expansions for the 
worldsheet fermions, we acquire 
\begin{eqnarray}
&~&\bigg{[}[\gamma(\delta_{\lambda}^{0}-v^{i}\delta_{\lambda}^{i})
-\gamma\mathcal{F}^{0}_{~~\bar\alpha}
\delta_{\lambda}^{\bar\alpha}]\psi^{\lambda}_{t}-
\nonumber\\
&-i\eta&[\gamma(\delta_{\lambda}^{0}-v^{i}\delta_{\lambda}^{i})
+\gamma\mathcal{F}^{0}_{~~\bar\alpha}
\delta_{\lambda}^{\bar\alpha}]\tilde{\psi}^\lambda_{-t}
\bigg{]}|B_\psi,\eta\rangle^{\rm T}=0~,
\nonumber\\
&~&\bigg{[}\left(\delta_{\lambda}^{\bar\alpha}
-\gamma^2\mathcal{F}_{~~0}^{\bar\alpha}
(\delta_{\lambda}^{0}-v^{i}\delta_{\lambda}^{i}) 
-\mathcal{F}_{~~\bar\beta}^{\bar\alpha}
\delta_{\lambda}^{\bar\beta}\right)\psi^{\lambda}_{t}-
\nonumber\\
&-i\eta&\left(\delta_{\lambda}^{\bar\alpha}+\gamma^2
\mathcal{F}_{~~0}^{\bar\alpha}
(\delta_{\lambda}^{0}-v^{i}\delta_{\lambda}^{i}) 
+\mathcal{F}_{~~\bar\beta}^{\bar\alpha}
\delta_{\lambda}^{\bar\beta}\right)\tilde{\psi}^\lambda_{-t}
\bigg{]}|B_\psi,\eta\rangle^{\rm T}=0~,
\nonumber\\
&~&\left[(\delta_\lambda^i-v^i\delta^0_\lambda)\psi_t^\lambda-i\eta
(-\delta_\lambda^i+v^i\delta^0_\lambda)
\tilde{\psi}^\lambda_{-t}\right]|B_\psi,\eta\rangle^{\rm T}=0~,
\nonumber\\
&~&\left[\psi_r^a +i\eta \tilde{\psi}^a_{-r}\right]
|B_\psi,\eta\rangle^{\rm T}=0~.
\end{eqnarray}
The boundary state equations of the usual 
(non-orbifoldy) directions,   
in the both twisted R-R and NS-NS sectors 
(for the nonzero-mode numbers of the R-R sector)
possess the following feature 
\begin{eqnarray}
\left(\psi_t^\lambda-i\eta S^\lambda_{(t)\lambda'}
\tilde{\psi}^{\lambda'}_{-t}\right)|B_\psi,\eta\rangle^{\rm osc ,T}=0~,
\end{eqnarray} 
where $t\in \mathbb{Z}-\{0\} \;(\mathbb{Z}+1/2)$ for the 
R-R (NS-NS) sector. The mode-dependent matrix 
$S^\lambda_{(t)\lambda'}$ is defined 
by $S_{(t)}=M_{(t)}^{-1}N_{(t)}$ with 
\begin{eqnarray}
M_{(t)\lambda}^{0}&=&\gamma(\delta_{\lambda}^{0}
-v^{i}\delta_{\lambda}^{i})-
\gamma\mathcal{F}^{0}_{~~\bar\alpha}\delta_{\lambda}^{\bar\alpha}~,
\nonumber\\
M_{(t)\lambda}^{\bar{\alpha}}&=&\delta_{\lambda}^{\bar{\alpha}}-
\gamma^2 \mathcal{F}^{\bar{\alpha}}_{~~0}(\delta_{\lambda}^{0}-v^{i}
\delta^{i}_{\lambda})
-\mathcal{F}^{\bar{\alpha}}_{~~\bar\beta}\delta^{\bar\beta}_{\lambda}~,
\nonumber\\
M_{(t)\lambda}^{i}&=&\delta_{\lambda}^{i}-v^{i}\delta_{\lambda}^{0}~,
\nonumber\\
\nonumber
N_{(t)\lambda}^{0}&=&\gamma(\delta_{\lambda}^{0}
-v^{i}\delta_{\lambda}^{i})+\gamma
\mathcal{F}^{0}_{~~\bar\alpha}\delta_{\lambda}^{\bar\alpha}~,
\nonumber\\
N_{(t)\lambda}^{\bar{\alpha}}&=&\delta_{\lambda}^{\bar{\alpha}}+\gamma^2
\mathcal{F}^{\bar{\alpha}}_{~~0}(\delta_{\lambda}^{0}-v^{i}
\delta^{i}_{\lambda})
+\mathcal{F}^{\bar{\alpha}}_{~~\bar\beta}\delta^{\bar\beta}_{\lambda}~,
\nonumber\\
N_{(t)\lambda}^{i}&=&-\delta_{\lambda}^{i}
+v^{i}\delta_{\lambda}^{0}~.\label{rty}
\end{eqnarray}
The boundary state equations of the orbifoldy directions, 
in the both twisted NS-NS and R-R sectors 
(for the nonzero-mode numbers of the NS-NS sector)
find the following form  
\begin{eqnarray}
\left(\psi_r^a+i\eta \tilde{\psi}^{a}_{-r}
\right)|B_\psi,\eta\rangle^{\rm osc ,T}=0~,
\end{eqnarray} 
where $r\in \mathbb{Z}+1/2 \;(\mathbb{Z}-\{0\})$ 
for the R-R (NS-NS) sector.

There are zero-mode parts in the both twisted NS-NS and R-R sectors. 
For the twisted NS-NS sector 
the zero-mode boundary state equation finds the feature 
\begin{equation}\left(\psi_0^a+i\eta
\tilde{\psi}^a_{0}\right){|B_\psi,\eta\rangle^{(0){\rm T}}_{\rm NS}}=0~.
\label{klop}
\end{equation}
Similarly, the zero-mode boundary state equation for 
the twisted R-R sector has the form  
\begin{eqnarray}
\left(\psi_0^\lambda-i\eta S^\lambda_{\;\;\;\lambda'}
\tilde{\psi}^{\lambda'}_{0}\right){|B_\psi,
\eta\rangle^{(0){\rm T}}_{\rm R}}=0~.\label{qw}
\end{eqnarray} 

\subsection{The solutions of the boundary state equations}

\subsubsection{The bosonic part}
We have calculated the bosonic part of the boundary state in the 
Ref. \cite{me}, hence we adduse the final feature of it. Its 
oscillating part is 
\begin{eqnarray}
|B_{X}\rangle^{\rm osc} 
&=&\prod_{n=1}^{\infty}[\det{M_{(n)}}]^{-1}\exp\left[
{-\sum_{m=1}^{\infty}
\left(\frac{1}{m}\alpha_{-m}^{\lambda}S_{(m)\lambda\lambda'}
\tilde{\alpha}_{-m}^{\lambda'}\right)}\right]\nonumber\\
\nonumber\\
&\times&\exp\left[-\sum_{r=1/2}^{\infty}
\left(\frac{1}{r}\alpha_{-r}^{a}\tilde{\alpha}_{-r}^{a}\right)\right]
|0\rangle_\alpha|0\rangle_{\tilde{\alpha}}~,\label{aos}
\end{eqnarray}
where the matrix $S_{(m)}$ is the same as one that 
was exhibited in the fermionic part.
The zero-mode portion of the bosonic boundary state is 
\begin{eqnarray}
|B_X\rangle^{(0)}=\frac{T_p}{2}
\int_{-\infty}^{\infty}\prod_{\lambda=0}^5
dp^{\lambda} \prod_i \left[
\delta \left(\hat{x}^{i}-v^{i}\hat{x}^{0}-y^{i}\right)
|p^{i}\rangle\right]\prod_{\alpha}|p^{\alpha}\rangle~
\label{zer}.
\end{eqnarray}
Note that since there is no motion along the orbifoldy 
directions, because the brane is stuck at the orbifold fixed-points 
$(x^a=0)$, Eq. (\ref{zer}) does not include 
bosonic zero modes for these directions.

\subsubsection{The NS-NS sector}

Using the coherent state method, the oscillating part of the 
boundary state in the twisted NS-NS sector possesses the form 
\begin{eqnarray}
|B_\psi,\eta\rangle^{\rm T}_{\rm NS}
&=&\prod_{s=1/2}^{\infty}[\det M_{(s)}]
~\exp\left[i\eta\sum_{t=1/2}^{\infty}\psi_{-t}^\lambda~ 
S_{(t)\lambda\lambda'}~\tilde{\psi}_{-t}^{\lambda'}\right]
\nonumber\\\nonumber\\
&\times& \exp\left[-i\eta\sum_{r=1}^{\infty}\psi_{-r}^a 
~\tilde{\psi}_{-r}^a\right]{|B_\psi,\eta
\rangle^{(0){\rm T}}_{\rm NS}}~,
\end{eqnarray}
where the zero-mode part of this sector 
comes from the orbifoldy directions, and 
according to Eq. (\ref{klop}) it is independent of 
the background fields and velocity. 
Therefore, it has the structure \cite{frac2}, \cite{frac3},
\begin{eqnarray}
{|B_\psi,\eta\rangle^{(0){\rm T}}_{\rm NS}}=\left(\bar{C}\;
\frac{1+i\eta\bar{\Gamma}}{1+i\eta}\right)_{LM}~
|L\rangle \otimes|\tilde{M}\rangle~,
\end{eqnarray}
where  $\bar{\Gamma}={\Gamma}^6 {\Gamma}^7
{\Gamma}^8{\Gamma}^9$, $\bar{C}$ is the 
charge conjugate matrix of $SO(4)$, and 
$|L\rangle$ and $|\tilde{M}\rangle$ are the spinors of $SO(4)$.

\subsubsection{The R-R sector}

Applying the same procedure of the bosonic part 
and the NS-NS sector, we acquire the following boundary 
state in the twisted R-R sector 
\begin{eqnarray}
|B_\psi,\eta\rangle^{\rm T}_{\rm R}
&=&\prod_{n=1}^{\infty}[\det M_{(n)}]
~\exp\left[i\eta\sum_{t=1}^{\infty}\psi_{-t}^\lambda~ 
S_{(t)\lambda\lambda'}~\tilde{\psi}_{-t}^{\lambda'}
\right]\nonumber\\\nonumber\\
&\times& \exp\left[-i\eta\sum_{r=1/2}^{\infty}
\psi_{-r}^a ~\tilde{\psi}_{-r}^a\right]
{|B_\psi,\eta\rangle^{(0){\rm T}}_{\rm R}}~,
\end{eqnarray}
in which ${|B_\psi,\eta\rangle^{(0){\rm T}}_{\rm R}}$ is 
the solution of Eq. (\ref{qw}). Its trusty form is 
given by 
\begin{eqnarray}
{|B_\psi,\eta\rangle^{(0){\rm T}}_{\rm R}}
=\left(\tilde{C}\;\Gamma^0\ldots\Gamma^p
\frac{1+i\eta\tilde{\Gamma}}{1+i\eta}~G\right)_{AB}~
|A\rangle \otimes|\tilde{B}\rangle~,
\end{eqnarray}
where $\tilde{\Gamma}={\Gamma}^0 {\Gamma}^1\ldots
{\Gamma}^5$, $|A\rangle$ and $|\tilde{B}\rangle$ are 
spinors of $SO(1,5)$, $\tilde{C}$ is the charge 
conjugate matrix of $SO(1,5)$, and the matrix 
$G$ is specified by the equation
\begin{eqnarray}\Gamma^\lambda G=S^\lambda _{\;\;\;\lambda '} 
G\Gamma^{\lambda '}.\end{eqnarray} 
The matrix $G$ has the conventional solution 
\begin{eqnarray}G=;\exp\left(\frac{1}{2}{\bar{\Phi}}_{\lambda\lambda'}
\Gamma^\lambda\Gamma^{\lambda'}\right);\;,\label{rt}
\end{eqnarray}
where the symbol $;~;$ means 
that we have to expand the exponential with the convention that
all $\Gamma$-matrices anticommute. This implies that
there are a finite number of terms in the expansion. Also
$\bar{\Phi}=(\Phi - \Phi^T)/2$, and  
the matrices $\Phi$ and $S$ have the relation
\begin{eqnarray}S=(\Lambda-\Phi)^{-1}(\Lambda'+\Phi),
\end{eqnarray}
in which  
\begin{eqnarray}\Lambda^0_\lambda =\gamma(\delta^0_\lambda 
-v^i\delta^i_\lambda )~~~&,&~~~\Lambda'^0_\lambda 
=\gamma(\delta^0_\lambda -v^i\delta^i_\lambda )~,\nonumber\\
\Lambda^{\bar{\alpha}}_\lambda =
\delta^{\bar{\alpha}}_\lambda ~~~~~~~~~~~~~~~&,&~~~
\Lambda'^{\bar{\alpha}}_\lambda =
\delta^{\bar{\alpha}}_\lambda ~,\nonumber\\
\Lambda^i_\lambda =\delta^i_\lambda ~~~~~~~~~~~~~~~&,&~~~~
\Lambda'^i_\lambda =-\delta^i_\lambda ~.
\end{eqnarray}
Thus, the matrix elements of $\Phi$ are given by 
\begin{eqnarray}
\Phi^0_\lambda &=&\gamma\mathcal{F}^{0}_{~~\bar\alpha}
\delta_{\lambda}^{\bar\alpha}~,\nonumber\\
\Phi^{\bar{\alpha}}_\lambda &=&\gamma^2
\mathcal{F}^{\bar{\alpha}}_{~~0}(\delta_{\lambda}^{0}-v^{i}
\delta^{i}_{\lambda})+\mathcal{F}^{\bar{\alpha}}_{~~\bar\beta}~
\delta^{\bar\beta}_{\lambda}~,\nonumber\\
\Phi^i_\lambda &=&v^i\delta^0_\lambda~.
\end{eqnarray}
Therefore, we find the 
following expressions for the matrix elements of $\bar{\Phi}$,
\begin{eqnarray}
\bar\Phi^0_\lambda &=&\frac{1}{2}
\left[(1-\gamma)\gamma\mathcal{F}^{0}_{~~\bar\alpha}
\delta_{\lambda}^{\bar\alpha}-v^i\delta^i_\lambda~\right],
\nonumber\\
\bar\Phi^{\bar{\alpha}}_\lambda &=&\frac{1}{2}\left[-(1-\gamma)
\gamma\mathcal{F}^{\bar{\alpha}}_{~~0}\delta_{\lambda}^{0}-\gamma^2 
\mathcal{F}^{\bar{\alpha}}_{~~0}v^{i}\delta^{i}_{\lambda}+2
\mathcal{F}^{\bar{\alpha}}_{~~\bar\beta}~
\delta^{\bar\beta}_{\lambda}~\right],
\nonumber\\
\bar\Phi^i_\lambda &=&\frac{1}{2}\left[v^i\delta^0_\lambda+\gamma^2 v^i 
\mathcal{F}_{~~\bar{\alpha}}^{0}\delta_{\lambda}^{\bar{\alpha}}\right]~.
\end{eqnarray}  
As an example, the explicit form of the matrix $\bar\Phi$ for
a fractional D2-brane that moves along the 
$x^3$-direction is
\begin{eqnarray}
\bar{\Phi}=\frac{1}{2}\left(
\begin{array}{cccccc}
0 &(1-\gamma)\gamma\mathcal{F}^{0}_{~~1} 
&(1-\gamma)\gamma\mathcal{F}^{0}_{~~2} 
&-v & 0 & 0\\
(\gamma-1)\gamma\mathcal{F}^{1}_{~~0}&0&
2\mathcal{F}^{1}_{~~2}&-v\gamma^2\mathcal{F}^{1}_{~~0} & 0 & 0\\
(\gamma-1)\gamma\mathcal{F}^{2}_{~~0}&2\mathcal{F}^{2}_{~~1}&0&
-v\gamma^2\mathcal{F}^{2}_{~~0} & 0 & 0\\
v&v\gamma^2\mathcal{F}^{1}_{~~0}
&v\gamma^2\mathcal{F}^{2}_{~~0}&0  & 0 & 0\\
0 & 0 & 0 & 0 & 0 & 0 \\
0 & 0 & 0 & 0 & 0 & 0 \\
\end{array}\right),
\end{eqnarray}
where $\mathcal{F}^{\bar \alpha}_{~~0}=\mathcal{F}^{0}_{~~{\bar \alpha}}$
and $\mathcal{F}^{\bar \alpha}_{~~{\bar \beta}}
=-\mathcal{F}^{\bar \beta}_{~~{\bar \alpha}}$.
%%%%%%%%%%%%%%%%%%%%%%%%%%%%%%%%%%%%%%%%%%%%%%%%%%%%%%%%%%%%%%%%%%
\section{Interaction of the D$p$-branes}
\hspace{0.4cm}
The interaction
between two D-branes can be described in two different
but equivalent ways, i.e. the open and closed string approaches. 
In the open string method the interaction amplitude 
is elaborated by the
one-loop diagram of an open string which is stretched between two
D-branes. In the
closed string approach, the interaction
of two D-branes is precisely recasted by the 
tree-level exchange of a closed
string that is emitted from one brane 
and is absorbed by another brane. 
We apply the second approach.

Now we are ready to calculate the interaction amplitude  
between two parallel fractional D$p$-branes 
with the velocities $v^i_{1}$ 
and $v^i_{2}$ in the twisted sector. 
For achieving this, we apply the interaction 
amplitude  
$\mathcal{A}_{\rm NS,R}^{\rm T}={}_{\rm NS,R}^{\rm T}\langle 
B_1|D|B_2\rangle_{\rm NS,R}^{\rm T}$, where 
$D$ is the closed string propagator 
$$D=2\alpha'\int_{0}^{\infty}dt~e^{-tH_{\rm closed}}~.$$
The closed string Hamiltonian $H_{\rm closed}$
is sum of the Hamiltonians of
the matter part and ghost part. 
The matter part of the Hamiltonian, in each of the NS-NS and R-R sectors, 
is sum of the Hamiltonians of
the bosonic part and the fermionic part, i.e.,
\begin{eqnarray}
H_{\rm Bos}&=&\alpha'p^{\lambda}p_{\lambda}
+2\sum_{n=1}^{\infty}(\alpha_{-n}^{\lambda}
\alpha_{n\lambda}
+\tilde{\alpha}_{-n}^{\lambda}\tilde{\alpha}_{n\lambda})
+2\sum_{r=1/2}^{\infty}
(\alpha_{-r}^{a}\alpha_{ra}
+\tilde{\alpha}_{-r}^{a}\tilde{\alpha}_{ra})-\frac{d-4}{6}~,
\nonumber\\
H_{\rm NS}&=&2\sum_{t=1/2}^{\infty}(t\psi_{-t}^{\lambda}
\psi_{t\lambda}
+t\tilde{\psi}_{-t}^{\lambda}\tilde{\psi}_{t\lambda})
+2\sum_{r=1}^{\infty}
(r\psi_{-r}^{a}\psi_{ra}
+r\tilde{\psi}_{-r}^{a}\tilde{\psi}_{ra})+\frac{d+2}{12}~, 
\nonumber\\
H_{\rm R}&=& 2\sum_{t=1}^{\infty}(t\psi_{-t}^{\lambda}
\psi_{t\lambda}
+t\tilde{\psi}_{-t}^{\lambda}\tilde{\psi}_{t\lambda})
+2\sum_{r=1/2}^{\infty}
(r\psi_{-r}^{a}\psi_{ra}
+r\tilde{\psi}_{-r}^{a}\tilde{\psi}_{ra})+\frac{d-4}{6}~.
\end{eqnarray}
$|B\rangle^{\rm T}_{\rm NS,R}$ 
is the total projected boundary state associated with the 
fractional D$p$-brane in the twisted sector
$$|B , \eta\rangle^{\rm T}_{\rm NS,R}=|B_{ \rm X}\rangle^{\rm T} 
\otimes|B_{\rm gh}\rangle\otimes|B_\psi,
\eta\rangle^{\rm T}_{NS,R}\otimes|B_{\rm sgh},\eta\rangle_{\rm NS,R}~,$$ 
where $|B_{\rm gh}\rangle$ and $|B_{\rm sgh},\eta\rangle_{\rm NS,R}$ 
are the ghost and superghost boundary states, respectively.  
The orbifold projection, the brane velocity and the background
fields do not change the (super)ghost boundary state. 

It is noticeable that a physical boundary state should
be invariant under the GSO-projection. Thus,
it is a linear combination of two states corresponding to 
$\eta=\pm 1$.
In the twisted sector, invariance of the boundary state  
under the action of the orbifold projection, in addition to 
the GSO-projection, changes the conventional form of the 
combination of the states, i.e. in the both twisted
NS-NS and R-R sectors there is a plus sign in the combination 
of the states \cite{bop}, \cite{bop2},
\begin{eqnarray}|B\rangle_{\rm NS,R}^{\rm T}=\frac{1}{2}
\left(|B,+\rangle_{\rm NS,R}^{\rm T}
+|B,-\rangle_{\rm NS,R}^{\rm T}\right)~.
\end{eqnarray}  

\subsection{The interaction amplitude in the NS-NS sector}

Using the total projected boundary states, the 
interaction amplitude in the twisted NS-NS sector finds 
the following feature 
\begin{eqnarray}
\mathcal{A}_{\rm NS-NS}^{\rm T}(\eta_1 , \eta_2)&=&\frac{2T_p^2
\alpha'V}{(2\pi)^{(5-p)/2}}
~\prod_{n=1}^{\infty}\frac{
\det \left(M_{(n-1/2)1} M_{(n-1/2)2}\right)}{
\det \left(M_{(n)1} M_{(n)2}\right)}
~\nonumber\\
\nonumber\\
&\times&\int_{0}^{\infty}dt(2\alpha' t)^{-5/2}~
\exp\left[ {-\frac{1}
{4\alpha't}\sum_{i}{\left(y_{1}^{i}-y_{2}^{i}\right)^2}}\right]
\nonumber\\
\nonumber\\ 
&\times & ~\delta_{\eta_1\eta_2,1}
\prod_{n=1}^{\infty}\frac{{\det\left[1+S^T_{(n-1/2)1}
S_{(n-1/2)2}q^{2n-1}\right]\left(1+q^{2n}\right)^4
\left(1-q^{2n}\right)^2 }}{{\det\left[1-S^T_{(n)1}S_{(n)2}q^{2n}
\right]~(1-q^{2n-1})^{4}\left(1+q^{2n-1}\right)^2}}~
~,\label{po}
\end{eqnarray}
where $q= e^{-2t}$, and $V$ is the common volume of the branes.
The damping exponential factor depends 
on the distance of the branes. The last line is induced 
by the oscillators of the matter 
part, conformal ghosts and 
super conformal ghosts. Since the ghost portion of 
the boundary state is not influenced 
by the velocities, background fields and orbifoldization,
the contribution of the ghosts 
has been introduced by manipulation of the amplitude. 

As it was mentioned, the 
orbifoldy directions induce the zero-mode boundary 
state in the twisted NS-NS sector. This 
brings in the factor `4' in the last line of the amplitude for 
$\delta_{\eta_1\eta_2,1}$, while its contribution
to $\delta_{\eta_1\eta_2,-1}$ is zero. 
Vanishing of this part of the interaction has significant 
effects on the amplitude of the distant branes, 
which will be discussed later.

\subsection{The interaction amplitude in the R-R sector}

Calculation of the interaction amplitude in the twisted R-R 
sector drastically needs more attention. 
This is due to the difficulties of the 
computations concerning to the divergent  
contribution of the zero-mode boundary state in this 
sector. Thus, 
we should use a suitable regularization to evaluate it.  
Similar to the Ref. \cite{lk}, we apply the regulator 
$x^{2(F_0+G_0)}$ as follows
\begin{eqnarray}
{}^{\rm T}{}_{\rm R}^{(0)}\langle B_1,\eta_1
|B_2,\eta_2\rangle_{\rm R}^{(0)}{}^{\rm T}
&\equiv& \lim_{x\rightarrow1}
\bigg{[}{}^{\rm T}{}_{\rm R}^{(0)}\langle B_{1\psi},
\eta_1|x^{2F_0}|B_{2\psi},
\eta_2\rangle_{\rm R}^{(0)}{}^{\rm T}
\nonumber\\
&\times& {}^{\rm T}{}_{\rm R}^{(0)}
\langle B_{1\rm sgh},\eta_1|x^{2G_0}|B_{2\rm sgh},
\eta_2\rangle_{\rm R}^{(0)}{}^{\rm T}\bigg{]}\nonumber,
\end{eqnarray}
where $(-1)^{F_0}=i\tilde{\Gamma}$ and $G_0=-\gamma_0\beta_0$.
The calculations of the superghost part eventuate to the same 
result of the untwisted sector
\begin{eqnarray}
{}^{\rm T}{}_{\rm R}^{(0)}\langle B_{1\rm sgh},
\eta_1|x^{2G_0}|B_{2\rm sgh},
\eta_2\rangle_{\rm R}^{(0)}{}^{\rm T}=\frac{1}{1+\eta_1\eta_2 x^2}~.
\end{eqnarray}
For the fermionic part we obtain 
\begin{eqnarray}
{}^{\rm T}{}_{\rm R}^{(0)}\langle B_{1\psi},\eta_1|x^{2F_0}|B_{2\psi},
\eta_2\rangle_{\rm R}^{(0)}{}^{\rm T}
&=&{\rm Tr} \left\{\bigg[\tilde{C}\;
\Gamma^0\ldots \Gamma^p 
\frac{1+i\eta_2\tilde{\Gamma}}{1+i\eta_2}
G_2\bigg]\tilde{C}^{-1}x^{2F_0}\right.\nonumber\\
&~&\left.\times\bigg{[}(-1)^p\tilde{C}\;\Gamma^0\ldots 
\Gamma^pG_1 \frac{1-i\eta_1\tilde{\Gamma}}{1+i\eta_1}
\bigg{]}^\dagger\tilde{C}^{-1}\right\}.
\end{eqnarray}
Substituting the explicit forms of $G_1$ and $G_2$ from Eq. (\ref{rt}),  
after a rigorous calculation, we receive the following regular factor  
for the interaction amplitude
\begin{eqnarray}
&~&{}^{\rm T}{}_{\rm R}^{(0)}\langle B_1,\eta_1|B_2,\eta_2
\rangle_{\rm R}^{(0)}{}^{\rm T}=
\nonumber\\
&~&\lim_{x\rightarrow1}
\bigg{\{}\left(\frac{1}{1+\eta_1\eta_2 x^2}\right) (-1)^p
\left(x+\frac{1}{x}\right)^3\delta_{\eta_1\eta_2,1}
\nonumber\\
&~&-\left(\frac{1}{1+\eta_1\eta_2 x^2}\right)
(-7)^{p+1}
\Psi_{\lambda_1\lambda_2}\Psi_{\lambda_3\lambda_4}
~\epsilon^{\lambda_1\lambda_2 \lambda_3\lambda_4}
\left(x-\frac{1}{x}\right)\left(x+\frac{1}{x}\right)^2~
\delta_{\eta_1\eta_2,-1}\bigg{\}},
\end{eqnarray}
where $\Psi_{\lambda_1\lambda_2}
=\frac{1}{2}(\bar{\Phi}_{(2)\lambda_1\lambda_2}
-\bar{\Phi}_{(1)\lambda_1\lambda_2})$ and 
$\epsilon^{\lambda_1\lambda_2 \lambda_3\lambda_4}$ 
is the Levi-Civita symbol. The factor $(-7)^{p+1}$ originates
from the anticommutation relations between 
the $\Gamma$-matrices which appeared in the product 
$\Gamma^0 \Gamma^1\ldots \Gamma^p$
and in the expansion of the exponential functions $G_1$ and $G_2$.
In the absence of the background fields the matrices 
$G_1$ and $G_2$ become trivial, and hence the 
conventional results of the literature will
be accurately reproduced. For acquiring Eq. (3.6) 
we have used the operators \cite{dive},
\begin{eqnarray}
N_1\equiv \Gamma^0\Gamma^1~~;~~N_2\equiv -i\Gamma^2
\Gamma^3~~;~~N_3\equiv -i\Gamma^4\Gamma^5~,\nonumber
\end{eqnarray}
which imply that the regulator of the fermionic 
zero modes satisfies the relations
\begin{eqnarray}
x^{2F_0}&=&x^{\sum_{k=1}^{3}N_k}~,\nonumber\\
{\rm Tr}[x^{2F_0}]&=&\prod_{k=1}^3{\rm Tr}[x^{N_k}]
=\left(x+\frac{1}{x}\right)^3~,\nonumber\\
{\rm Tr}[x^{2F_0}\Gamma^0\Gamma^1\Gamma^2\Gamma^3]
&=&i{\rm Tr}[x^{N_1}N_1]\;
{\rm Tr}[x^{N_2}N_2]\;{\rm Tr}[x^{N_3}]=i\left(x-\frac{1}{x}\right)^2
\left(x+\frac{1}{x}\right).~~~~~~~~~~
\end{eqnarray}

Finally, the interaction amplitude in the twisted R-R sector is
\begin{eqnarray}
\mathcal{A}_{\rm R-R}^{\rm T}(\eta_1 , \eta_2)&=&
\frac{2T_p^2\alpha'V}{(2\pi)^{(5-p)/2}}
\int_{0}^{\infty}dt(2\alpha' t)^{-5/2}~
\exp\left( {-\frac{1}
{4\alpha't}\sum_{i}{\left(y_{1}^{i}-y_{2}^{i}\right)^2}}\right)
\nonumber\\
\nonumber\\
&\times &\left\{
(-1)^p\delta_{\eta_1\eta_2,1}
~\prod_{n=1}^{\infty}\frac{\det\left[1+S^T_{(n)1}
S_{(n)2}q^{2n}\right]
\left(1+q^{2n-1}\right)^4 \left(1-q^{2n}\right)^2
}{\det\left[1-S^T_{(n)1}S_{(n)2}q^{2n}\right]~
(1-q^{2n-1})^{4}\left(1+q^{2n}\right)^2}~
\right.
\nonumber\\
\nonumber\\ 
&-&\left.
(-7)^{p+1}
\Psi_{\lambda_1\lambda_2}\Psi_{\lambda_3\lambda_4}
~\epsilon^{\lambda_1\lambda_2 \lambda_3\lambda_4}~
\delta_{\eta_1\eta_2,-1}\right\}.\label{oi}
\end{eqnarray} 

The total amplitude in the twisted sector is 
\begin{eqnarray}
\mathcal{A}_{\rm total}^{\rm T}
=\mathcal{A}^{\rm T}(+ , +)
+\mathcal{A}^{\rm T}(- , -)
+\mathcal{A}^{\rm T}(+ , -)
+\mathcal{A}^{\rm T}(- , +)~.
\end{eqnarray}
where 
$\mathcal{A}^{\rm T}(\eta_1 , \eta_2)
=\mathcal{A}_{\rm NS-NS}^{\rm T}(\eta_1 , \eta_2)
+\mathcal{A}_{\rm R-R}^{\rm T}(\eta_1 , \eta_2)$. 
Comparing the resulted amplitudes (\ref{po}) and (\ref{oi}) with 
the conventional forms of the interaction amplitudes
reveals that the presence of the orbifoldy directions 
induces significant effects on the interaction.

Note that the complete form of the theory contains  
both the utwisted and twisted sectors. 
That is, the total interaction 
amplitude is sum of the amplitude of the twisted sector,
i.e. Eq. (3.9), and the amplitude in the 
untwisted sector. In fact, since for the various 
setups the untwisted sector has been deeply studied 
we only investigated the twisted sector of 
our setup.
%%%%%%%%%%%%%%%%%%%%%%%%%%%%%%%%%%%%%%%%%%%%%%%%%%%%%%%%%%%%%
\subsection{Interaction of the distant branes}

For obtaining the long-range forces of the theory, the 
behavior of the interaction amplitude for distant 
branes should be evaluated. 
Hence, we discuss the 
evolution of the interaction amplitude under 
the limit $t\rightarrow \infty$
for the oscillating part of the amplitudes in the both twisted sectors.
The limit is not applied on the first and second lines of 
Eqs. (\ref{po}) and (\ref{oi}), because these factors 
originate from the bosonic zero modes, and the oscillators do not 
have any contribution to them.
We shall see that the orbifold projection, 
accompanied by the background fields, 
imposes some new effects on the 
long-range forces. 

The interaction amplitude for the twisted NS-NS sector, 
after passing a long enough time, is
\begin{eqnarray}
{\tilde {\mathcal{A}}}^{\rm T}_{\rm NS-NS}(\eta_1 , \eta_2)&=&
\frac{2T_p^2\alpha'V}{(2\pi)^{(5-p)/2}}
~\prod_{n=1}^{\infty}\frac{
\det \left(M_{(n-1/2)1} M_{(n-1/2)2}\right)}{
\det \left(M_{(n)1} M_{(n)2}\right)}
~\nonumber\\
\nonumber\\
&\times&\int_{0}^{\infty}dt(2\alpha' t)^{-5/2}
\exp\left( {-\frac{1}
{4\alpha't}\sum_{i}{\left(y_{1}^{i}-y_{2}^{i}\right)^2}}\right)
\nonumber\\
\nonumber\\
&\times &\delta_{\eta_1\eta_2,1}
~\lim_{t\rightarrow\infty}\left[1+\left({\rm Tr}(S^T_{(1/2)1}
S_{(1/2)2})+2\right)e^{-2t}\right].
\end{eqnarray}
This amplitude contains three terms. We observe 
that the sum of the second and third terms, which is 
induced by the massless 
states (i.e. graviton, dilaton and Kalb-Ramond), ghosts 
and superghosts, vanishes. This is due to the fact that the 
orbifold projection deformed the zero-point energy of the
closed string 
Hamiltonian, therefore, this peculiar result was appeared.
However, even under this condition, the first term,
i.e. ``1'' in the last line, remains. 
Note that in the untwisted sector the first term, i.e. ``1''
disappears, and the second and third terms,
without the damping factor $e^{-2t}$, withstand.

The twisted R-R sector possesses the following contribution 
to the amplitude of the distant branes 
\begin{eqnarray}
{\tilde {\mathcal{A}}}^{\rm T}_{\rm R-R}(\eta_1 , \eta_2)&=&
\frac{2T_p^2\alpha'V}{(2\pi)^{(5-p)/2}}
~\int_{0}^{\infty}dt(2\alpha' t)^{-5/2}
\exp\left( {-\frac{1}
{4\alpha't}\sum_{i}{\left(y_{1}^{i}-y_{2}^{i}\right)^2}}\right)
\nonumber\\
\nonumber\\ 
&\times & ~\left[
(-1)^p\delta_{\eta_1\eta_2,1}
-(-7)^{p+1}
\Psi_{\lambda_1 \lambda_2}\Psi_{\lambda_3 \lambda_4}
~\epsilon^{\lambda_1 \lambda_2
\lambda_3 \lambda_4}~\delta_{\eta_1\eta_2,-1}\right] ~,
\end{eqnarray}
where the last
line originates from the fermionic zero-mode part. 

The total amplitude in the twisted sector, 
due to the exchange of the massless states, is 
\begin{eqnarray}
{\tilde {\mathcal{A}}}_{\rm total}^{\rm T}
={\tilde {\mathcal{A}}}^{\rm T}(+ , +)
+{\tilde {\mathcal{A}}}^{\rm T}(- , -)
+{\tilde {\mathcal{A}}}^{\rm T}(+ , -)
+{\tilde {\mathcal{A}}}^{\rm T}(- , +)~.
\end{eqnarray}
where 
${\tilde {\mathcal{A}}}^{\rm T}(\eta_1 , \eta_2)
={\tilde {\mathcal{A}}}_{\rm NS-NS}^{\rm T}(\eta_1 , \eta_2)
+{\tilde {\mathcal{A}}}_{\rm R-R}^{\rm T}(\eta_1 , \eta_2)$.

Again note that the complete form of the theory includes  
both the twisted and untwisted sectors. 
Therefore, the total amplitude of the long-range force 
is sum of the amplitude of the twisted sector, i.e. Eq. (3.12),
and that one of the untwisted sector. 
%%%%%%%%%%%%%%%%%%%%%%%%%%%%%%%%%%%%%%%%%%%%%%%%%%%%%%%%%%
\section{Conclusions}

As a first result, for a fractional moving D$p$-brane
with internal background fields which 
lives in an orbifoldized 
spacetime $\mathbb{R}^{1, 5} \times\mathbb{C}^{2}/
\mathbb{Z}_{2}$, we constructed a corresponding boundary state.
The background fields are 
Kalb-Ramond tensor and a $U(1)$ gauge potential. The 
orbifoldy directions divide the boundary state into 
two different parts: the twisted and untwisted sectors
\cite{frac2}. 
Unlike the untwisted sector, the boundary state 
of the twisted NS-NS sector includes a zero-mode portion.
Our setup imposed a modified 
zero-mode part for the twisted R-R sector.

As another result, the interaction 
amplitude of two parallel moving 
fractional D$p$-branes, or a system of one fractional brane and 
one fractional anti-brane with the background fields,
for the both twisted NS-NS and R-R sectors, was acquired. 
The orbifoldy directions introduced some 
significant effects on the interaction. 
For example, in the amplitude of the twisted NS-NS sector,  
the contribution of the GSO-projected sectors with 
different spin structures vanishes. However,
various parameters in the amplitude, 
i.e. the parameters of the background fields, the velocities,
the branes dimensions, accompanied by the
orbifoldy directions, elegantly give a generalized 
feature to this interaction. Thus, the strength 
of the interaction accurately is adjustable.

Obtaining the interaction amplitude for distant 
branes is the other conclusion. 
We observed that in the both twisted sectors 
contribution of the massless states, unlike the conventional case, 
prominently vanishes. In addition, in the twisted NS-NS sector 
we received an extra term. These are effects 
of the orbifoldization.

{\bf Added Notes:}

By setting all background fields and the 
velocities to zero our results reduce to the simple setups of
the D-branes, e.g. see the Refs. \cite{dive, lk, pez}.

According to the paper \cite{cra} and the following 
references therein [5, 6, 7, 65], 
the orbifold conformal field theory is obtained with
a non-trivial background $B$-field. This implies that 
the Cardy's condition for our setup, i.e., that in the
presence of the background fields and the velocities 
which have resemblance with the electric components of the
anti-symmetric $B$-field, is satisfied.

%$$$$$$$$$$$$$$$$$$$$$$$$$$$$$$$$$$$$$$$$$$$$$$$$$$$$$$$$$$$$$$$$


\begin{thebibliography}{99}
\bibitem{pol0}
J. Polchinski, 
Phys. Rev. Lett. \textbf{75} (1995) 4724.
\bibitem{bon} 
E. Cremmer and J. Scherk, Nucl. Phys. \textbf{B
50}(1972) 222.
\bibitem{bon2} 
L. Clavelli and  J. Shapiro,  
Nucl. Phys. \textbf{B 57} (1973) 490.
\bibitem{hus} 
F. Hussain, R. Iengo and  C. Nunez, Nucl. Phys.
\textbf{B 497} (1997) 205.
\bibitem{bergf}
O. Bergman, M. Gaberdiel and  G. Lifschytz, 
Nucl. Phys. \textbf{B 509} (1998) 194.
\bibitem{dex1}
P. Di Vecchia, M. Frau, I. Pesando, S. Sciuto, A. Lerda and
R. Russo, Nucl. Phys. \textbf{B 507} (1997) 259.
\bibitem{dex2}
M. Billo, P. Di Vecchia and D. Cangemi, Phys. Lett.
\textbf{B 400} (1997) 63.
\bibitem{fra} 
M. Frau, I. Pesando, S. Sciuto, A. Lerda and  R. 
Russo, Phys. Lett. \textbf{B 400} 
(1997) 52.
\bibitem{calli} 
C. G. Callan and  I. R. Klebanov, Nucl. Phys. 
\textbf{B 465} (1996) 473.
\bibitem{guk} 
S. Gukov, I. R. Klebanov and  A. M. Polyakov, 
Phys. Lett. \textbf{B 423} (1998) 64.
\bibitem{li}
M. Li, Nucl. Phys. \textbf{B 460} (1996) 351.
\bibitem{kit} 
T. Kitao, N. Ohta and  J. G. Zhou, Phys. Lett. 
\textbf{B 428} (1998) 68.
\bibitem{bach0}
C. Bachas, Phys. Lett. \textbf{B 374} (1996) 37.
\bibitem{dex4} 
E. T. Akhmedov, M. Laidlaw and G. W. Semenoff, JETP Lett. 
\textbf{77} (2003) 1-6; M. Laidlaw and 
G. W. Semenoff, JHEP 0311:021, (2003).

\bibitem{doc1}
H. Arfaei and D. Kamani, 
Phys. Lett. \textbf{B 452} (1999)54, hep-th/9909167;
Nucl. Phys. \textbf{B 561} (1999) 57-76, hep-th/9911146;
Phys. Lett. \textbf{B 475} (2000) 39-45, hep-th/9909079;
D. Kamani, Phys. Lett. \textbf{B 487} (2000) 187-191,
hep-th/0010019; Nucl. Phys. {\bf B 601} (2001) 149, arXiv:hep-th/0104089; 
Phys. Lett. \textbf{B 580} (2003) 257-264, hep-th/0301003;
Int. J. Theor. Phys. \textbf{47} (2008) 1533-1541, hep-th/0611339;
Eur. Phys. J. \textbf{C 26} (2002) 285-291, hep-th/0008020;
Mod. Phys. Lett. \textbf{A 17} (2002) 237-244, hep-th/0107184;
Annals of Physics \textbf{354} (2015) 394-400, 
arXiv:1501.02453[hep-th];
F. Safarzadeh-Maleki and D. Kamani,
Phys. Rev. \textbf{D 90}, 107902 (2014), arXiv:1410.4948[hep-th];
Phys. Rev. \textbf{D 89}, 026006 (2014), arXiv:1312.5489[hep-th]. 
E. Maghsoodi and D. Kamani, 
{\it Fractional-wrapped branes with rotation, linear motion
and background fields}, Nucl. Phys. B (2017), 
http://dx.doi.org/10.1016/j.nuclphysb.2017.07.009 .

\bibitem{frac0} 
M. R. Douglas, JHEP \textbf{9707} (1997) 004.
\bibitem{dog1} 
D. Diaconescu, M. R. Douglas and  J. Gomis,  
JHEP \textbf{9802} (1998) 013.
\bibitem{frac1} 
C. V. Johnson and  R. C. Myers, Phys. Rev.  
\textbf{D 55} (1997) 6382-6393.
\bibitem{frac2} 
M. Frau, A. Liccardo and  R. Musto,  Nucl. 
Phys. \textbf{B 602} (2001) 39-60.
\bibitem{frac3} 
M. Bertolini, P. Di Vecchia, M. Frau, A. Lerda and  R.
Marotta,  Nucl. Phys. \textbf{B 621} (2002) 157.
\bibitem{dive}
P. Di Vecchia, A. Liccardo, R. Marotta and  F. Pezzella,
JHEP \textbf{0306} (2003) 007.

\bibitem{me}
M. Saidy-Sarjubi and D. Kamani, 
Phys. Rev. D \textbf{92}, (2015) 046003, arXiv:1508.02084[hep-th].

\bibitem{bop}
M. R. Gaberdiel and  J. B. Stefanski, Nucl. Phys. 
\textbf{B 578} (2000) 58-84.
\bibitem{bop2}
I. Kriz, L. A. Pando Zayas and  N. Quiroz, Int. J. 
Mod. Phys. \textbf{A 23} (2008) 933-974.
\bibitem{lk}
M. Billo, P. Di Vecchia, M. Frau, A. Lerda, I. 
Pesando, R. Russo and  S. Sciuto, Nucl. Phys. 
\textbf{B 526} (1998) 199-228.

\bibitem{pez}
P. Di Vecchia, A. Liccardo, R. Marotta and F. Pezzella, 
Int. J. Mod. Phys. \textbf{A 20} (2005) 4699-4796.

\bibitem{cra}
M. Billo, B. Craps and F. Roose, JHEP \textbf{0101} (2001) 038.
%$$$$$$$$$$$$$$$$$$$$$$$$$$$$$$$$$$$$$$$$$$$$$$$$$$$$$$$$$$$$$$$$$$$$$$$$$$$$
\end{thebibliography}
\end{document}